\documentclass[11pt]{article}

\usepackage[body={16cm,230mm}]{geometry}
\usepackage{epsfig,amsfonts,amssymb,amsmath}
\usepackage{amsthm}
\usepackage{enumerate}
\usepackage{cite}
\def\a{\alpha}
\def\b{\beta}
\def\g{\gamma}

\def\l{\lambda}

\def\s{\sigma}

\def\beq{\begin{equation}}
\def\eeq{\end{equation}}
\def\be{\begin{displaymath}}
\def\ee{\end{displaymath}}
\def\bea{\begin{eqnarray}}
\def\eea{\end{eqnarray}}

\def\bmat{\left(\begin{array}}

\def\ds{\displaystyle}

\def\Wp{ \raise.4ex\hbox{\textrm{\Large $\wp$}}}

\def\eqref#1{(\ref{#1})}
\def\?{(?)\marginpar{|?}}

\renewcommand{\author}[1]{\large\rm #1\\ \bigskip}
\newcommand{\address}[1]{{\normalsize\it #1\\}\bigskip}
\renewcommand{\title}[1]{\bigskip\bigskip\Large\bf #1\bigskip\bigskip\\}

\newcounter{app}
\newcounter{sapp}[app]
\def\theapp{\Alph{app}}
\newcommand{\app}[1]{
\refstepcounter{app}{\vspace{7mm}
\noindent\Large\bf Appendix
\theapp.
 \ #1 \par \vspace{5mm}}
\setcounter{equation}{0}
\def\theequation{\Alph{app}.\arabic{equation}}}

\begin{document}

%\begin{flushright}
%{\sf version 1.0 \it (5/10/2004)}
%\end{flushright}

\vglue .3 cm

\begin{center}
\title{ Form factor expansions in the 2D Ising model
      and  Painlev\'e VI}

\author{         Vladimir V. Mangazeev\footnote[1]{email:
                {\tt Vladimir.Mangazeev@anu.edu.au}}}
\address{Department of Theoretical Physics,\\
         Research School of Physics and Engineering,\\
    Australian National University, Canberra, ACT 0200, Australia}
\author{              Anthony J. Guttmann\footnote[2]{email:
                {\tt tonyg@ms.unimelb.edu.au}}}
\address{ ARC Centre of Excellence for\\
Mathematics and Statistics of Complex Systems,\\
Department of Mathematics and Statistics,\\
The University of Melbourne, Victoria 3010, Australia}

\end{center}

\setcounter{footnote}{0}
\vspace{5mm}

\begin{abstract}
We derive a Toda-type recurrence relation, in both high and
low temperature regimes, for the $\lambda$ - extended
diagonal correlation functions $C(N,N;\lambda)$ of the
two-dimensional Ising model, using an earlier connection
between diagonal form factor expansions and tau-functions
within Painlev\'e VI (PVI) theory, originally discovered
by Jimbo and Miwa. This greatly simplifies
the calculation of the diagonal correlation functions,
particularly their $\lambda$-extended counterparts.

We also conjecture a closed form expression for the simplest
off-diagonal case $C^{\pm}(0,1;\lambda)$ where a connection
to PVI theory is not known.
Combined with the results for diagonal correlations
these give all the
initial conditions required for the
$\l$-extended version of
quadratic difference equations for the correlation
functions discovered by McCoy, Perk and Wu. The results
obtained here should provide a further
potential algorithmic improvement in the
$\l$-extended case, and facilitate other developments.

\end{abstract}

\newpage
%%%%%%%%%%%%%%%%%%%%%%%%%%%%%%%%%%%%%%%%%%%%%%%%%%
\section{Introduction}\label{sect:intro}

The 2D Ising model in zero magnetic field is arguably
the most important solvable
model in theoretical statistical mechanics.
Its free energy \cite{O44} and the spontaneous magnetization
\cite{O49,MR0051740} were computed a long time ago
by Onsager and Yang.

Despite more than half a century of attempts by many
outstanding scientists, a closed form expression for
the susceptibility is still unknown.
We have a vast amount of numerical and analytical
information, mostly derived from careful analysis
of exact series expansions. The best algorithm
for such series expansions, due to Orrick, Nickel, 
Guttmann and Perk \cite{ONGP00}, is based
on expressing the susceptibility in terms of correlation
functions, and using quadratic
difference equations, due to McCoy, Perk and Wu 
\cite{McW80}, \cite{MPW81}, \cite{Perk80} to recursively calculate
the required correlation functions. A related study
that gives complementary analytic information is based
on the multi-particle expansion~\cite{WuMcTr76} of
the susceptibility derived by Wu, McCoy Tracy and
Barouch in 1976. In either formulation, one
requires the correlation functions or
their $\lambda$- extended counterparts (see (\ref{temp3}-\ref{temp4}) below).

In this paper we present a different method for the
calculation of
diagonal pair correlation functions and form factor
expansions that occur in the theory of the 2D
Ising model in zero magnetic field.

We shall restrict ourselves
 to the case of the symmetric 2D Ising model
 defined by the interaction energy
\beq
 E=-J\sum_{i,j}(\s_{i,j}\s_{i,j+1}+\s_{i,j}\s_{i+1,j}), \label{intro1}
\eeq
where $J>0$ (ferromagnetic regime) and spins $\s_{i,j}=\pm1$ are
assigned to each site of a square lattice.

The 2D Ising model has a second order phase transition
at the critical temperature $T_c$ defined by the condition
\beq
\sinh(2J/k_BT_c)=1,\label{intro2}
\eeq
where $k_B$ is Boltzmann's constant.

A study of the pair correlation functions
\beq
C(M,N)=\langle\sigma_{0,0}\>\sigma_{M,N}\rangle  \label{intro3}
\eeq
started long ago  \cite{KO1949} when  they were expressed
in terms of determinants \cite{Montroll63,McCoyWu}.
Subsequent studies of correlation functions
\cite{Wu66,Kadan1966,ChengWu,WuMcTr76,BaxCor1978,Perk80,Ghosh84,Martinez97}
led to infinite  form factor expansions with
every form factor given by a multiple
integral \cite{WuMcTr76}.

Note also that there are two special cases where a simpler
determinant representation
for correlation functions is known. These
are the cases of row and
diagonal correlations $M=0$ and $M=N$ respectively,
for any $N$. In both cases there exists a
representation for pair correlation functions by
$N\times N$ Toeplitz determinants
\cite{McCoyWu}.

In 1980 Jimbo and Miwa \cite{JM1980} found yet another
approach to the calculation
of the diagonal correlations $C(N,N)$. They showed
that $C(N,N)$ satisfies the
``sigma'' form of the Painlev\'e VI equation and produced a system of
recurrence relations which allows one to calculate diagonal correlations
$C(N+1,N+1)$ in terms of $C(N,N)$ and  $C^*(N,N)$
which are the correlation functions  obtained by
the duality transformation interchanging
high/low temperature regimes.

Moreover, Jimbo and Miwa introduced the so-called $\l$-extension
(or isomonodromic deformation) of the correlation functions using
a multi-particle expansions in terms of free fermion operators.
Then they
related this to the matrix  Schlesinger equations with integrability conditions
given by the Painlev\'e VI equation.\footnote{The idea of a 
$\l$-extension was made even earlier in McCoy, Tracy and 
Wu \cite{MTW77} in connection with solutions to Painlev\'e III.}
In such an approach the parameter $\l$ describes monodromy properties of
solutions for Schlesinger equations and plays the role of boundary conditions
for the Painlev\'e VI equation.

Similar $\l$-extensions naturally appear in the
Fredholm determinant approach \cite{WuMcTr76} and can be
regarded as generating functions for the form factors
(see (\ref{temp3}-\ref{temp4}) below).
The case $\l=1$ corresponds to the symmetric Ising model.

Starting with the asymptotics  of diagonal correlations
in the high/low temperature regimes, the
authors of \cite{Holon} studied  series solutions of the
Painlev\'e VI equation
 and showed the existence of the one-parameter extension
 of $C(N,N)$ which
exactly matches the form factor extension.
They conjectured that the Jimbo-Miwa $\l$-extension and
the form factor $\l$-extension
are actually the same, with the parameter $\l$ in (\ref{temp3}-\ref{temp4}) playing
the role of boundary condition for the Painlev\'e VI equation.

We note that such an identification is a result of experimental
mathematics, confirmed by a large amount of  computer calculation.
In this paper we will rely on this, even though
a rigorous mathematical proof of such an identification is still missing.

A different approach to calculation of correlation functions
was developed
by McCoy and Wu \cite{McW80} and by Perk \cite{Perk80} in 1980. 
They  discovered a set of quadratic difference equations
satisfied by the pair correlation functions of the 2D Ising model.
These equations were developed further in \cite{McCoyWu81}.

In the symmetric
case the quadratic difference equations allow one to calculate all 
correlation functions recursively,
once we know the ``initial conditions'':
diagonal correlations $C(N,N)$ and $C^*(N,N)$ together
with the nearest-neighbor correlations $C(0,1)$ and $C^*(0,1)$ \cite{Perk02}.
The diagonal correlations can be  calculated from the
Jimbo-Miwa recurrence relations and when combined with well known expressions
for the nearest-neighbor correlations, gives
a very effective method for series calculations of the correlation functions and
the susceptibility \cite{ONGP00}.

In this paper we present a different method of
calculation of diagonal correlations in the $\l$-extended case. It is based
on the explicit relationship of such  correlations to tau-functions
in Painlev\'e VI theory. We also conjecture a formula for the $\l$-extension
of the nearest neighbor correlations $C(0,1)$. The series in the elliptic nome
for the form factors contributing to $C(0,1)$ were first conjectured
in sec.~5.2 of \cite{ONGP00}. At that time it was not clear how to sum  up
such expressions\footnote{After this paper appeared, the technique
for summing up such series was given by McCoy et al. in \cite{Mccoy2010}.
}.
Here we have used a completely different method to obtain a closed form expression
for $C(0,1)$ in terms of elliptic functions.

Following \cite{PF06} we will use two variables $s$ and $t;$
\beq
t=s^4,\quad s=\begin{cases}
 \sinh(2E/k_B T)^{\phantom{-1}},\> & T>T_c,\\
{\sinh(2E/k_B T)}^{-1},\> & T<T_c.
\end{cases}
\label{intro4}
\eeq
Let us introduce a special notation for diagonal correlation
functions
\beq
C^\pm_N(t)=C(N,N),\label{intro5}
\eeq
where $\pm$ denotes $T\gtrless T_c$
regimes, respectively. We note in the symmetric case the dual
of $C^\pm_N(t)$ is $C^\mp_N(t)$ .

Our main result is the following Toda-type relation
for diagonal correlations
\beq
t\frac{d^2}{dt^2}
\log C^\pm_N(t)+\frac{d}{dt}\log C^\pm_N(t)
+\frac{N^2}{(1-t)^2}=
\frac{\left(N^2-\frac{1}{4}\right)}{(1-t)^2}\,
\frac{C^\pm_{N+1}(t)\,C^\pm_{N-1}(t)}
{(C^\pm_N(t))^2},\label{intro6}
\eeq
valid for both temperature regimes at $N=1,2,\ldots$.
Supplemented by initial conditions for $N=0,1,$ relation
(\ref{intro6}) gives a very effective way to calculate $C^\pm_N(t)$ 
for $N=2,3,4,\ldots$. It took us under 10 mins
to calculate $C^+_N(t)$ for $N=2,\ldots,20$ on a laptop
computer using a Mathematica program.

What's more important is that  eqn (\ref{intro6}) remains
valid for the $\l$-extension of diagonal correlation functions.
In fact, as shown in section 5,
a form factor expansion for $N=0$ is sufficient to
calculate the generalized diagonal correlations for $N=1,2,\ldots$.

The paper is organized as follows.
In section 2 we define the $\l$ - extended pair
correlation functions, give integral formulae for the form factors
and recall the result of Jimbo and Miwa on diagonal correlators.
In section 3 we list some
basic facts on the Hamiltonian formulation of Painlev\'e VI theory
and prove a Toda-type  relation for tau-functions
related by a special shift transformation. In section 4
we establish a connection with the Jimbo-Miwa result and prove
relation (\ref{intro6}). In section 5 we calculate
generalized diagonal correlators at arbitrary $\l$
for $N=0,1,2$.
In section 6 we present a closed form expression for the
$\l$ - extension of the first nontrivial off-diagonal
correlation function $M=0$, $N=1$. Together with the results of Section 4,
these provide all initial conditions required for
the quadratic difference equation satisfied by
the $\l$-extended correlation functions,
and so allows, in principle, the rapid calculation of any required
correlation function.
The major advance is in the calculation of
$\lambda$-extended correlation functions.
In section 7 we discuss the results obtained,
and mention possible further developments.

%%%%%%%%%%%%%%%%%%%%%%%%%%%%%%%%%%%%%%%%%%%%%%%%%%
\section{Form factor expansions of pair correlation functions}\label{corfun}
%%%%%%%%%%%%%%%%%%%%%%%%%%%%%%%%%%%%%%%%%%%%%%%%%%

We start with the
$\lambda$ - expansion of pair correlation functions
in multi-particle components
\beq
C^+(M,N;\lambda)=s^{-1}(1-t)^{1/4}\sum_{n=0}^\infty
\lambda^{2n+1}\hat C^{(2n+1)}(M,N),\quad T>T_c, \label{temp3}
\eeq
and
\beq
C^-(M,N;\lambda)=(1-t)^{1/4}(1+\sum_{n=1}^\infty
\lambda^{2n}\hat C^{(2n)}(M,N)),\quad T<T_c,   \label{temp4}
\eeq
where for $n>0$ (see, for example, formula (4.2) in \cite{ONGP00},
given here for the symmetric case)
\beq
\hat C^{(n)}(M,N)=\frac{1}{n!}\hspace{-0.2cm}
\mathop{\int}_{-\pi_{\phantom{-}}}^{\phantom{++}
\pi_{\phantom{,}}}\frac{d\omega_1}{2\pi}\ldots\hspace{-0.2cm}
\mathop{\int}_{-\pi_{\phantom{-}}}^{\phantom{++}
\pi_{\phantom{,}}}\frac{d\omega_n}{2\pi}
\left[\prod_{i=1}^n\frac{x_i^M}{\sinh\gamma_i}\right]
\Biggl[\,\prod_{1\leq i<j\leq n}h_{ij}\Biggr]^2
\,\cos\Bigl(N\sum_{i=1}^n\omega_i\Bigr),               \label{temp5}
\eeq
\beq
\sinh\gamma_i=(y_i^2-1)^{1/2},\quad x_i=y_i-(y_i^2-1)^{1/2},\quad
y_i=s+s^{-1}-\cos\omega_i, \label{temp6}
\eeq
\beq
h_{ij}=\frac{2(x_i x_j)^{1/2}\sin((\omega_i-\omega_j)/2)}{1-x_ix_j}.
\label{temp7}
\eeq

We shall call $C^\pm(M,N;\l)$ {\it generalized
correlation functions.}
The pair correlations for the symmetric Ising model
are obtained by setting
$\lambda=1$
\beq
C^\pm(M,N)=C^\pm(M,N;1).   \label{temp8}
\eeq

Using a different approach Jimbo and Miwa \cite{JM1980} in 1980
introduced the function
\beq
\sigma_N(t)=\begin{cases}
{\ds t(t-1){\frac{d}{dt}}_{\phantom{I}}\hspace{-0.2cm}
\log C^+(N,N)-\frac{1}{4}}, & T>T_c,\\
{\ds t(t-1){\frac{d}{dt}}^{\phantom{I}}\hspace{-0.2cm}
\log C^-(N,N)-{\frac{t}{4}}}, & T<T_c.
\end{cases}  \label{temp9}
\eeq
and  showed that
it satisfies the ``sigma'' form
of Painlev\'e VI
\beq
\left(t(t-1)\frac{d^2\s}{dt^2}\right)^2+
4\frac{d\sigma}{dt}\left((t-1)\frac{d\sigma}{dt}-\s-\frac{1}{4}\right)
\left(t\frac{d\sigma}{dt}-\s\right)=
N^2\left((t-1)\frac{d\sigma}{dt}-\s\right)^2.  \label{temp10}
\eeq
In 2004 \cite{FW04} Forrester and Witte gave a new derivation of this result
working from the Toeplitz determinant form rather
than the Fredholm form.

As mentioned in the introduction, Jimbo and Miwa also considered
an isomonodromic $\l$-extension of
$C^\pm(N,N)$ which also satisfies (\ref{temp10}).
Using extensive computer calculations Boukraa et al. \cite{Holon}
observed that the Jimbo-Miwa $\l$-extension is the same
as the form factor expansions (\ref{temp3}-\ref{temp4}).
 It provides an opportunity
to investigate properties of generalized diagonal correlation functions
$C^\pm(N,N;\l)$ by studying solutions of (\ref{temp10}).

In the rest of the paper we will assume
$\lambda$ to be arbitrary  unless its value is explicitly specified.
We shall also use the same notation
introduced in (\ref{intro5}) for
generalized diagonal correlation functions (\ref{temp3}-\ref{temp4})
\beq
C^\pm_N(t)\equiv C^\pm(N,N;\lambda)   \label{temp11}
\eeq
and simply omit the dependence on $\lambda$.

The solutions of (\ref{temp10}) can be  specified by their expansion
near $t\to0$.
In \cite{Holon} Boukraa et al. analyzed series expansions of
$C^\pm_N(t)$ at $t\to 0$ and demonstrated that
\beq
C^-_N(t)=(1-t)^{1/4}+
\lambda^2\frac{(1/2)_N(3/2)_N}{4((N+1)!)^2}t^{N+1}(1+O(t))
\label{temp12}
\eeq
and
\beq
C^+_N(t)=\lambda(1-t)^{1/4}
f^{(1)}_{N,N}+
\l^3\frac{(1/2)_N((3/2)_N)^2}{16(N+1)!(N+2)!}t^{3N/2+2}(1+O(t)),\label{temp13}
\eeq
where
\beq
f^{(1)}_{N,N}=t^{N/2}\frac{(1/2)_N}{N!}\phantom{|}_2F_1
\left(\frac{1}{2},N+\frac{1}{2},N+1;t\right),\label{temp13a}
\eeq
and $(a)_N = \Gamma(a+N)/\Gamma(a)$ is the usual Pochhammer symbol.
We shall
use (\ref{temp12}-\ref{temp13})  to fix a normalization of
recurrence relations in section 4.

\section{Some facts from Painlev\'e VI theory}

In this section we shall briefly review the  properties of the
Hamiltonian form of the Painlev\'e VI (${\bf P}_{VI}$) equation
\cite{Pain02,Gamb10}, given by eqn (\ref{temp16}).
Our main reference is the work of Okamoto \cite{Ok87}.

For each solution of the ${\bf P}_{VI}$ equation, one can construct a
Hamiltonian function $H(t)\equiv H(t;{\bf{b}})$ which depends on
four parameters
\beq
{\bf b}=(b_1,b_2,b_3,b_4).\label{temp14}
\eeq
A tau-function $\tau(t)\equiv\tau(t;{\bf{b}})$ related to the Hamiltonian
$H(t;{\bf{b}})$ is defined by
\beq
H(t;{\bf{b}})=\frac{d}{dt}\log\tau(t;{\bf{b}}).\label{temp14a}
\eeq

Let us introduce an auxiliary Hamiltonian $h(t)\equiv h(t;{\bf b})$,
\beq
h(t)=t(t-1)H(t)+e_2(b_1,b_3,b_4)
\,t-\frac{1}{2}e_2(b_1,b_2,b_3,b_4)\label{temp15}
\eeq
where $e_i(x_1,\ldots,x_n)$ is the $i$-th elementary symmetric
function in $n$ variables
and a set of $x_i$'s can be a subset of $b_i$'s as in (\ref{temp15}).

The function $h(t)$ solves
the equation ${\bf E}_{{VI}}[{\bf b}]$\footnote{So named by
Okamoto~\cite{Ok87} in his studies of PVI.
It is in fact a generalization of the  ``sigma" form of PVI
(\ref{temp10}).}
\beq
h'(t)\Bigl[t(1-t)h''(t)\Bigr]^2
+\Bigl[h'(t)[2h(t)-(2t-1)h'(t)]+b_1b_2b_3b_4\Bigr]^2=
\prod_{k=1}^4\Bigl(h'(t)+b_k^2\Bigr)\ .\label{temp16}
\eeq

The group $G$ of Backlund transformations of ${\bf P}_{VI}$ is
isomorphic to the affine Weyl group of type $F_4$: $W_a(F_4)$. It contains
the following transformations of parameters
(not all of them are independent)
\beq
w_1: b_1\leftrightarrow b_2,\quad w_2:
b_2\leftrightarrow b_3,\quad w_3: b_3\leftrightarrow b_4,
\quad w_4: b_3\to -b_3,\> b_4\to-b_4,\label{temp17}
\eeq
\beq
x^1: \kappa_0\leftrightarrow\kappa_1,\quad
x^2: \kappa_0\leftrightarrow\kappa_\infty,\quad
x^3: \kappa_0\leftrightarrow\theta\label{temp18}
\eeq
where
\beq
\kappa_0=b_1+b_2,\quad \kappa_1=b_1-b_2,
\quad \kappa_\infty=b_3-b_4,\quad \theta=b_3+b_4+1\label{temp19}
\eeq
and the parallel transformation
\beq
l_3: {\bf b}\equiv(b_1,b_2,b_3,b_4)
\to{\bf b^+}\equiv(b_1,b_2,b_3+1,b_4).\label{temp20}
\eeq

For each transformation $s$ from the group $W_a(F_4)$ such that
\beq
s: {\bf b}\mapsto {\bf b_s}\label{temp21}
\eeq
one can construct another auxiliary Hamiltonian $h(t;{\bf b_s})$
which satisfies ${\bf E}_{{VI}}[{\bf b_s}]$.
The function $h(t;{\bf b_s})$ is a rational function of $t$ and
\beq
\left \{h(t;{\bf b}),\>\frac{d}{dt}h(t;{\bf b}),\>\frac{d^2}{dt^2}h(t;{\bf b})
\right\}.\label{temp22}
\eeq
If we consider a sequence of transformations related to $l_3^m$,
$m=0,\pm1,\pm2,\ldots$ and construct an associated sequence
$h_m(t)$ and related tau-functions $\tau_m(t)$, then it is well known
\cite{Ok87}
that such tau-functions satisfy Toda-type recurrence relations.

Let us consider a different transformation $l_s\in G$:
\beq
l_s=x_1\,x_2\,l_3\,x_2\,x_1,  \label{temp23}
\eeq
where the action of the $x_i$'s and $l_3$ on $b_i$'s is defined by
(\ref{temp17}-\ref{temp20}).

After simple calculations one obtains
\beq
l_s^{\pm1}: (b_1,b_2,b_3,b_4)\mapsto (b_1\pm\frac{1}{2},b_2\mp\frac{1}{2},
b_3\pm\frac{1}{2},b_4\pm\frac{1}{2}). \label{temp24}
\eeq

To calculate the expressions for related auxiliary Hamiltonians
we need to use the results from Okamoto's paper \cite{Ok87}, sections 2,3.
Here we shall only present the answer

\beq
h_\pm(t)\equiv h(t,{\bf b^\pm})=
h(t)-\frac{1}{4}\pm\frac{\tilde S_1}{4}+
\frac{2t(t-1)^2h''(t)\pm[2\tilde S_1h(t)+
2\tilde S_3(1-t)-\tilde S_1\tilde S_2]}
{2(2(t-1)h'(t)-2h(t)+\tilde S_2)}\label{temp25}
\eeq
where

\beq
\tilde S_i=e_i(b_1,-b_2,b_3,b_4),
\quad {\bf b^\pm}\equiv l_s^{\pm1}({\bf{b}})=
(b_1\pm\frac{1}{2},b_2\mp\frac{1}{2},
b_3\pm\frac{1}{2},b_4\pm\frac{1}{2}).\label{temp26}
\eeq

Now we can follow the Okamoto derivation of
Toda relations for tau-functions.
Using (\ref{temp15}) and (\ref{temp25}) we obtain
\bea
&&h_+(t)+h_-(t)-2h(t)=
t(t-1)\left[H_+(t)+H_-(t)-2H(t)+\frac{3t}{2}\right]=\nonumber\\
&&=-\frac{t}{2}+\frac{t(t-1)^2h''(t)}{2(t-1)h'(t)-
2h(t)+\tilde S_2}.\label{temp27}
\eea
Substituting
\beq
H_\pm(t)=\frac{d}{dt}\log[\tau_\pm(t)],\quad
H(t)=\frac{d}{dt}\log[\tau(t)]\label{temp28}
\eeq
and integrating over $t$ we get
\beq
B\,(t-1)^2\frac{\tau_+(t)\,\tau_-(t)}{\tau^2(t)}=
(t-1)h'(t)-h(t)+\frac{\tilde S_2}{2},\label{temp29}
\eeq
or after using the expression for $h(t)$ in terms of $\tau(t),$
\beq
t\frac{d^2}{dt^2}
\log\tau(t)+\frac{d}{dt}\log\tau(t)=
B\,\frac{\tau_+(t)\,\tau_-(t)}{\tau^2(t)},
\label{temp30}
\eeq
where $B$ is the integration constant.

We have just shown that if we introduce a sequence of parameters
\beq
{\bf b}_N\equiv \left(b_1+\frac{N}{2},b_2-\frac{N}{2},b_3+\frac{N}{2},
b_4+\frac{N}{2}\right)\label{temp31}
\eeq
and associated sequences of Hamiltonians and tau-functions
\beq
h_N(t)=h(t,{\bf b}_N),\quad H_N(t)=H(t,{\bf b}_N),\quad
\tau_N(t)=\tau(t,{\bf b}_N),\label{temp32}
\eeq
then tau-functions $\tau_N(t)$ satisfy Toda-type recurrence relations
\beq
t\frac{d^2}{dt^2}
\log\tau_N(t)+\frac{d}{dt}\log\tau_N(t)=
B_N\,\frac{\tau_{N+1}(t)\,\tau_{N-1}(t)}{\tau^2(t)},\label{temp33}
\eeq
where the normalization constant  $B_N$ can
depend on the parameters $b_1,b_2,b_3,b_4$
and $N$.

\section{Recurrence relations for $C_N^\pm(t)$}

Now we use the results from the previous section to obtain recurrence
relations for generalized correlation functions $C_N^\pm(t)$ valid
for arbitrary $\l$.

First we need to establish a connection between equations (\ref{temp10})
and (\ref{temp16}). Let us consider the equation
${\bf E}_{{VI}}[{\bf b}_N]$ (\ref{temp16}) with parameters
(\ref{temp31})
and choose
\beq
b_1=-1/2,\quad b_2=-1/2,\quad b_3=0,\quad b_4=0.\label{temp35}
\eeq
If we denote a corresponding solution of (\ref{temp16}) as $h_N(t)$, then
after a substitution
\beq
h_N(t)=\s_N(t)-\frac{N^2}{4}\,t+\frac{1}{8},\label{temp36}
\eeq
 equation (\ref{temp16}) coincides with (\ref{temp10}).

Now combining formulae (\ref{temp9}, \ref{temp14a}-\ref{temp15}) and
(\ref{temp36})
we can find a connection between tau-functions $\tau_N(t)$ from the previous
section and $C^\pm_N(t),$ viz.
\beq
\tau_N(t)=\begin{cases}
{\ds C^+_N(t)(1-t)^{-N^2}t^{\frac{N}{2}}}, & T>T_c\\
{\ds C^-_N(t)(1-t)^{-N^2}t^{\frac{N}{2}-\frac{1}{4}}}, & T<T_c.
\end{cases}\label{temp37}
\eeq
Substituting (\ref{temp37}) into (\ref{temp33}) we finally
arrive at recurrence relations
for generalized diagonal correlation functions $C^\pm_N(t),$
\beq
t\frac{d^2}{dt^2}
\log C^\pm_N(t)+\frac{d}{dt}\log C^\pm_N(t)
+\frac{N^2}{(1-t)^2}=
\frac{\left(N^2-\frac{1}{4}\right)}{(1-t)^2}\,
\frac{C^\pm_{N+1}(t)\,C^\pm_{N-1}(t)}
{(C^\pm_N(t))^2}.\label{temp38}
\eeq
The normalization constant $B_N$ in (\ref{temp33}) can be
calculated by substituting  expansions (\ref{temp12}-\ref{temp13}) at $t\to0$
into (\ref{temp38}). It appears that
\beq
B_N=N^2-\frac{1}{4}\label{temp39}
\eeq
for both the $T>T_c$ and $T<T_c$ regimes.

Let us comment on these  Toda-type relations for $C^\pm_N(t)$
which are valid for arbitrary values of $\l$.
They allow us to calculate diagonal correlations for any $N=2,3,4,\ldots\>$
starting with $C_0(t)$ and $C_1(t)$.
We don't need to use
correlations at the dual temperature $s^{-1}$ as in \cite{JM1980}.
In fact, when $\lambda$ is arbitrary,
it is sufficient to know  $C_0^\pm(t)$ only, as
$C_1^\pm(t)$ can then be calculated from (\ref{temp38})
as we shall see in the next section.

To our knowledge, equations (\ref{temp38}) have not previously appeared
in the literature. They provide a very efficient method for the
 calculation of diagonal form-factors
and  sums in formulae (\ref{temp3}-\ref{temp4}) in closed form.

\section{The Picard solution and calculation of $C_0^\pm(t)$ and $C_1^\pm(t)$}

In this section we shall calculate initial conditions for the
recursion (\ref{temp38}). The functions $C_0^\pm(t)$ are simply related to
tau-functions $\tau_0(t)$ by (\ref{temp37}).
A related function $h_0(t),$ given by eqn (\ref{temp36}) solves the
equation ${\bf E}_{{VI}}[{\bf b}]$ (\ref{temp16}) with parameters $b_i$ given by eqn
(\ref{temp35}).
It is well known that a general solution of (\ref{temp16})
depending on two arbitrary parameters (specified by initial conditions)
can be found in this case.

Consider another set of parameters $b_i$
\beq
b_1=0,\quad b_2=0,\quad b_3=-\frac{1}{2},\quad b_4=-\frac{1}{2}.\label{temp40}
\eeq
This choice corresponds to the Picard solution \cite{Pic89}
of the Painlev\'e VI equation
which can be found in closed form (see, for example, \cite{Maz01})
and expressed in terms of the elliptic Weierstrass function depending on
two arbitrary parameters.

One can represent solutions 
of Painlev\'e equations
in terms of logarithmic derivatives
of ratios of functions with no movable singularities.
As one of many examples of such a representation, we mention
a representation of solutions of the ${\bf P}_{VI}$ as logarithmic derivatives
of tau-functions (see section 4.3 of \cite{Ok87}).

A location of zeros of tau-functions for the Picard solution
(and related the Hitchin solution)
has been investigated by Kitaev and Korotkin \cite{Kit98} and  
recently by Brezhnev \cite{Brezh09}.
It follows that the tau-function for the Picard case should be proportional to the
Jacobi theta-function depending on parameters specified by initial conditions.
A detailed calculation of tau-functions for the Picard class solutions has been
given recently by the first author in \cite{Mang2010}. Here we will use these results.

First, let us denote an auxiliary Hamiltonian and related tau-function
for the Picard case (\ref{temp40}) as $h_{P}(t)$ and $\tau_{P}(t)$ respectively.
Since the equation (\ref{temp16}) is symmetric in the $b_i$'s we immediately
conclude that
\beq
h_0(t)=h_{P}(t). \label{temp41}
\eeq

Using (\ref{temp15}) and integrating over $t$ we produce
a relation between tau-functions
\beq
\tau_0(t)=(1-t)^{1/4}\tau_{P}(t).\label{temp42}
\eeq

We introduce an elliptic parameterization and define
the elliptic modulus $k$ as
\beq
k=s^2,\quad t=k^2. \label{temp43}
\eeq

Following \cite{WW} we will define  theta functions
$\theta_i(x|\tau)$ with quasi-periods $\pi$ and $\pi\tau$.
The elliptic modulus  $k$ in (\ref{temp43}) and its complement $k'$ are
\beq
k=\frac{\theta_2(0|\tau)^2}{\theta_3(0|\tau)^2},\quad
k'=\frac{\theta_4(0|\tau)^2}{\theta_3(0|\tau)^2},\quad k^2+k'^2=1.\label{temp44}
\eeq
It will be more convenient to use the parameter $t$ as the second argument of the
elliptic functions
\beq
\tau=i\frac{\mathrm{K}'(t)}{\mathrm{K}(t)},\quad q=e^{i\pi\tau},
\quad \mathrm{K}(t)=\frac{\pi}{2}\theta_3^2(0|\tau),\quad \mathrm{K}'(t)=K(1-t),
\label{temp45}
\eeq
where
${\mathrm{K}}(t)$ and ${ \mathrm{K}}'(t)$
are the complete elliptic integrals of the first kind of
parameters $t$ and $1-t$ respectively.

Now let us introduce Jacobi elliptic functions
\beq
\mbox{sn}(u,t)=\frac{1}{k^{1/2}}
\frac{\theta_1(v|\tau)}{\theta_4(v|\tau)},\quad
\mbox{cn}(u,t)=\frac{(k')^{1/2}}{k^{1/2}}
\frac{\theta_2(v|\tau)}{\theta_4(v|\tau)},\quad
\mbox{dn}(u,t)={(k')^{1/2}}
\frac{\theta_3(v|\tau)}{\theta_4(v|\tau)},\label{temp46}
\eeq
where
\beq
u=\frac{2\mathrm{K}(t)}{\pi}v\label{temp47}
\eeq
and define the fundamental elliptic integral of
the second kind \cite{WW} as
\beq
\mathcal{E}(u,t)=\int_0^u\mbox{dn}^2(x,t)dx.\label{pain25}
\eeq
It satisfies
\beq
\mathcal{E}(\mathrm K(t),t)=\mathrm{E}(t),\label{temp48}
\eeq
where $\mathrm E(t)$ is the complete elliptic integral of the second kind.

It is convenient to use slightly renormalized complete elliptic integrals
\beq
K=\frac{2\mathrm{K}(t)}{\pi},\quad E=\frac{2\mathrm{E}(t)}{\pi}.\label{temp48a}
\eeq

One of the results of \cite{Mang2010} can be written as
 \beq
 \tau_P(t)=c_0(x,y)\,{q^{{y^2/{\pi^2}}}}{t^{-1/4}}\,
 \frac{\theta_1(x+\tau y|\tau)}
{\theta_4(0|\tau)},\label{temp49}
\eeq
where $x$ and $y$ are two arbitrary parameters
and $c_0(x,y)$ is a normalization
constant independent of $t$.

First let us consider the high temperature regime $T>T_c$.
Combining (\ref{temp37}, \ref{temp42}, \ref{temp49}) we obtain:
\beq
C_0^+(t)=c_0(x,y)\frac{(1-t)^{1/4}}{t^{1/4}}q^{y^2/\pi^2}
\frac{\theta_1(x+\tau y|\tau)}{\theta_4(0|\tau)}\label{temp50}
\eeq
for some values of $x$ and $y$. Recall that $C_0^\pm(t)$ depends
on the parameter $\l$. It remains to find
a relationship between the parameters $x$ and $y$ on the RHS and
the parameter $\l$ on the LHS of (\ref{temp50}).
We can expand both parts of (\ref{temp50}) in series in  $x,$ $y$ and $\l$
and compare the expansions term by term.

Integrating the well known formula for
a logarithmic derivative of $\theta_4(x|\tau)$
we obtain
\beq
\theta_4(x|\tau)=\theta_4(0|\tau)
\exp\Bigl\{-\frac{x^2}{2} K E+
\sideset{}{}\int_0^{\>\>\>xK}\mathcal {E}(y,t)dy\Bigr\}. \label{temp51}
\eeq

Using (\ref{temp46}) and (\ref{temp51}) we can expand the rhs of
(\ref{temp50}) in a series in $x$ and $y\tau$ with polynomial coefficients
depending on $t$, $K$ and $E$.

For $C_0^+(t)$ we need to calculate the first
few form factors in the sum (\ref{temp3}).
To that end, we can use the integral representation (\ref{temp5}) or a different
integral representation \cite{Holon} for the functions
\beq
f_{N,N}^{(2n+1)}={\hat C^{(2n+1)}(N,N)}/{s}, \quad
f_{N,N}^{(2n)}={\hat C^{(2n)}(N,N)}. \label{temp52}
\eeq
It was observed in \cite{Holon} that the diagonal
form factors
$f_{N,N}^{(n)}$ are polynomials in $t$, $K$ and $E$
for small values of
$n,$ and the first four, corresponding to $n=1 \ldots 4$
are given in ~\cite{Holon}. Now we can expand the integral representation
for the diagonal form-factors
in series in $t$ up to sufficiently high order and compare
it with the above polynomial Ansatz. In that way one
can produce closed form expressions for the form-factors.
The first
few exact form factors $\hat C^{(n)}(0,0)$, $n=1,2,\ldots,7$
are given in
Appendix A. We note that in \cite{ONGP00}, eqn (5.7), Orrick et al.
gave a conjectured expression for $\hat C^{(n)}(0,0)$
giving the results in terms of $q$ series.
Evaluation of their expression for $n = 1,2,\ldots,7$
gives a series expansion that agrees with the expansion
of the expressions given in Appendix A, thus providing mutual conformation.

Now matching series on both sides of (\ref{temp50}) up to order
$\lambda^8$ we achieve perfect agreement if we choose
\beq
c_0(x,y)=1,\quad
\lambda=x-\frac{x^3}{6}+\frac{x^5}{120}-\frac{x^7}{5040}+O(x^9),\quad
y=0.   \label{temp53}
\eeq
From this expansion we can guess that
\beq
\l=\sin (x)\label{temp54}
\eeq
and we obtain
\beq
C^+_0(t)=C^+(0,0;\l)=\frac{(1-t)^{1/4}}{t^{1/4}}
\frac{\theta_1(\arcsin(\l)|\tau)}{\theta_4(0|\tau)},\label{temp55}
\eeq
a result which has not been published before.
A nontrivial self-check of the relation (\ref{temp54})
is to set $\l=1$ which corresponds to the symmetric Ising model.

We obtain
\beq
C^+(0,0;1)=\frac{(1-t)^{1/4}}{t^{1/4}}
\frac{\theta_1(\frac{\pi}{2}|\tau)}{\theta_4(0|\tau)}=
\frac{(1-t)^{1/4}}{t^{1/4}}
\frac{\theta_2(0|\tau)}{\theta_4(0|\tau)}=1\label{temp56}
\eeq
as expected.

Another comment is in order. Based on connections to  Painlev\'e VI theory
we have {\bf proved} that the generalized diagonal correlations
can be represented
in terms of theta-functions. However, the relations between the
parameters $x,y$ and $\l$
are still conjectural. One of the possible ways to prove
such relations
is to find a partial differential equation for generalized correlations
which includes derivatives with respect to {\bf both} $\l$ and $t$.

We can handle the low-temperature regime $T<T_c$  in a similar way.
The only difference is that
now we have different relations between parameters,
\beq
c_0(x,y)=-i\exp(ix),\quad \l=\sin(x), \quad y=\pi/2. \label{temp57}
\eeq
The answer
reads
\beq
C_0^-(t)=C^-(0,0;\l)=
(1-t)^{1/4}\frac{\theta_4(\arcsin(\l) |\tau)}{\theta_4(0|\tau)}=
\frac{\theta_4(\arcsin(\l) |\tau)}{\theta_3(0|\tau)}
.\label{temp58}
\eeq
Again for $\l=1$
\beq
C_0^-(t)=C^-(0,0;1)=\frac{\theta_4(\frac{\pi}{2} |\tau)}{\theta_3(0|\tau)}=1.
\label{temp58a}
\eeq

The formula (\ref{temp58}) first appeared  in Boukraa et al.\cite{Holon}
(see formula (92), Section 6). The authors of \cite{Holon}
remark that they have calculated $\l$-extended diagonal
correlations for
$N=0,1,2,\ldots$, and have given a selection of the algebraic 
curves which arise from special choices of $\l.$\footnote
{After this paper appeared, results for $N=0,1$
diagonal correlations were
given in \cite{Mccoy2010}, which follow from conjectured results in \cite{ONGP00}.}

We emphasize that in our derivation of
(\ref{temp55}, \ref{temp58}) we relied
on the exact calculation of tau-functions for
the Picard solution \cite{Mang2010}.
This provides an alternative derivation of equation (\ref{temp58}),
previously announced in \cite{Holon}.

As mentioned above we have conjectured relations between
parameters (\ref{temp53}-\ref{temp54})
and (\ref{temp57})   based on series
expansions in $\l$ and $x$
and the Ising case $\l=1$.
However, formulae (\ref{temp56}) and (\ref{temp58})
produce analytic expressions
for the integrals (\ref{temp5}) at $M=N=0$ for
any positive integer $n$.
We  checked numerically up to $n=10$ that we
obtain correct values for the
form factors in (\ref{temp5}).

Now we turn to the calculation of $C^\pm_1(t)$.
Let us start with the auxiliary Hamiltonian $h_0(t)$
corresponding to the
Picard solution. We take the tau-function (\ref{temp49})
with
$y=0$ ($y=\pi/2$) and arbitrary $x$ for $T>T_c$ ($T<T_c$)
and use eqns.
(\ref{temp14a}-\ref{temp15}) with parameters (\ref{temp35})
to find $h_0(t)$.
Now we can find auxiliary Hamiltonians $h_\pm(t)$
corresponding to $N=\pm1$
from (\ref{temp25}-\ref{temp26}).
It is easy to see that they coincide
\beq
h_+(t)=h_-(t).  \label{temp59}
\eeq
It follows from (\ref{temp14a}-\ref{temp15})
and (\ref{temp31}) with $N=\pm1$ that
\beq
\tau_{1}(t)=a(x)\,t\,\tau_{-1}(t) \label{temp60}
\eeq
and from (\ref{temp37})
\beq
C_1^\pm(t)=a_\pm(x)C_{-1}^\pm(t)\label{temp61}
\eeq
for some integration constants $a_\pm(x)$.

Substituting into (\ref{temp38}) at $N=0$ we obtain
\beq
({C_{1}^\pm(t)})^2=
-\frac{4(1-t)^2}{a_\pm(x)}(C_0^\pm(t))^2\left\{t\frac{d^2}{dt^2}
\log C^\pm_0(t)+\frac{d}{dt}\log C^\pm_0(t)\right\}.\label{temp62}
\eeq
We can calculate first and second derivatives of $C_0^\pm(t)$.
For that we need the following
differentiation formulae

\beq
\frac{d}{dt}K=\frac{E}{2t(1-t)}-\frac{K}{2t},\quad
\frac{d}{dt}E=\frac{E-K}{2t},\label{temp63}
\eeq
\beq
\frac{d}{d t}\mbox{sn}(u,t)=
-\frac{\mbox{sn}(u,t)\mbox{cn}^2(u,t)}{2(t-1)}+
\frac{\mbox{cn}(u,t)\mbox{dn}(u,t)}{2t(t-1)}\bigl[u(t-1)+
\mathcal {E}(u,t)\bigr],\label{temp64}
\eeq
\beq
\frac{d}{d t}\mbox{cn}(u,t)=\phantom{-}
\frac{\mbox{sn}^2(u,t)\mbox{cn}(u,t)}{2(t-1)}-
\frac{\mbox{sn}(u,t)\mbox{dn}(u,t)}{2t(t-1)}\bigl[u(t-1)+
\mathcal{E}(u,t)\bigr],\label{temp65}
\eeq
\beq
\frac{d}{d t}\mbox{dn}(u,t)=\phantom{-}
\frac{\mbox{sn}^2(u,t)\mbox{dn}(u,t)}{2(t-1)}-
\frac{\mbox{sn}(u,t)\mbox{cn}(u,t)}{2(t-1)}\bigl[u(t-1)+
\mathcal {E}(u,t)\bigr],\label{temp66}
\eeq
\beq
\frac{d}{d t}\mathcal {E}(u,t)=
\frac{\mathcal {E}^2(u,t)}{2(t-1)}-
\frac{u}{2}\mbox{sn}^2(u,t)-\frac{\mbox{sn}(u,t)
\mbox{cn}(u,t)\mbox{dn}(u,t)}
{2(t-1)}\label{temp66a}
\eeq
and
\beq
\frac{d}{d
t}\left[\log\left(\frac{\theta_4(x|\tau)}
{\theta_4(0|\tau)}\right)\right]=
\frac{\mbox{sn}^2(xK,t)}{4(1-t)}+
\frac{(\mathcal {E}(xK,t)-x E)^2}{4t(t-1)}.\label{temp66b}
\eeq
The last formula can be proved using (\ref{temp51}) and
a partial differential equation for
theta-functions.

After straightforward simplifications we arrive at
\beq
t\frac{d^2}{dt^2}
\log C^-_0(t)+\frac{d}{dt}\log C^-_0(t)=
-\frac{1}{4(1-t)^2}\left({\mbox{cn}(z,t)\,\mbox{dn}(z,t)}
+{\mbox{sn}(z,t)}\,X\right)^2
\label{temp67}
\eeq
and
\beq
t\frac{d^2}{dt^2}
\log C^+_0(t)+\frac{d}{dt}\log C^+_0(t)=
-\frac{1}{4t(1-t)^2}
\frac{X^2}{\mbox{sn}^2(z,t)}
\label{temp68}
\eeq
where
\beq
z=xK=\frac{2x\mathrm K(t)}{\pi},
\quad
X=\frac{1}{K}\left[\log\theta_4(x|\tau)\right]'_x=\mathcal{E}(xK,t)-xE.
\label{temp69}
\eeq
Substituting back into (\ref{temp62}) we have
\beq
C_1^-(t)=\frac{1}{\sqrt{a_-(x)}}
\frac{\theta_4(x|\tau)}{\theta_3(0|\tau)}
\left({\mbox{cn}(z,t)\mbox{dn}(z,t)}
+{\mbox{sn}(z,t)}\,X\right)
\label{temp70}
\eeq
and
\beq
C_1^+(t)=\frac{1}{\sqrt{a_+(x)}}\,
\frac{\theta_4(x|\tau)}{\theta_3(0|\tau)}\,\frac{X}
{t^{1/2}}. \label{temp71}
\eeq
To determine the factors $a_\pm(x)$ we use (\ref{temp54}),
expand (\ref{temp70}-\ref{temp71}) in $\l$ and
compare with form factor
expansions (\ref{temp3}-\ref{temp4}) at $M=N=1$.
These form-factors are given
in \cite{Holon} up to order nine in $\l$ and we
will not reproduce
them here. All these expansions are consistent with a choice
\beq
\sqrt{a_+(x)}=\sqrt{a_-(x)}=\cos(x).\label{temp72}
\eeq

Again we can do an independent check by expanding
near the point $x=\pi/2$ or $\l=1$ which
corresponds to the symmetric
Ising model. In this case we obtain the results for
the nearest-neighbor diagonal correlations $C(1,1)$
\beq
C^+(1,1;1)=t^{-1/2}\left[{(t-1)K+E}\right]\label{temp73}
\eeq
and
\beq
C^-(1,1;1)=E,\label{temp74}
\eeq
which agree with formulae (4.9a) and (4.9b), Section VIII.4 of \cite{McCoyWu}.

Now we can substitute the initial values of
$C^\pm_0(t)$, $C^\pm_1(t)$ into recurrence
relation (\ref{temp38}) and calculate $C^\pm_N(t)$
for any $N=2,3,\ldots$ for
arbitrary $\l$. All we need are
the differentiation formulae given above.

We obtain for $N=2$ and $T<T_c$
\beq
C^-_2(t)=\frac{\theta_4(x|\tau)}{\cos^2(x)\,\theta_3(0|\tau)}
\Biggl[C^2+\frac{8}{3}
\,C\,S\,X+X^2\left(2S^2-\frac{t+1}{3t}\right)
-\frac{X^4}{3t}\Biggr]
\label{temp74a}
\eeq
and for $T>T_c$
\bea
&&C^+_2(t)=-\frac{\theta_4(x|\tau)}{3\,t\,\cos^2(x)\,\theta_3(0|\tau)}
\biggl[2\,C\,X\left[2\,X^2+2\,t\,S^2-t-1\right]+
\nonumber\\
&&\phantom{C^-_2(t)}
+S\left[X^4+X^2(6\,t\,S^2-5(t+1)\right]+t\,C^2\biggr]\label{temp74b}
\eea
where  $X$ is defined in (\ref{temp69}), and
\beq
C=\mbox{cn}(z,t)\,\mbox{dn}(z,t),\quad S=\mbox{sn}(z,t).\label{temp74c}
\eeq
The answer for $C_3^\pm(t)$ is given in Appendix B.

Let us point out that the factor $C_{N-1}^\pm(t)$
{\it always} divides the lhs of (\ref{temp38}).
We don't have a rigorous
proof of this cancelation. Exactly the same
phenomenon occurs for many polynomial
tau-functions related to Painlev\'e equations.
Usually the proof of this fact relies on
determinant representations of such tau-functions.
It would be extremely interesting to find a determinant representation
of $C_N^\pm(t)$ for arbitrary $\l$.

As mentioned in the Introduction,
 the equations (\ref{temp38})  for $\l=1$ together
with  initial conditions
(\ref{temp56}, \ref{temp58a}, \ref{temp73}-\ref{temp74})
and differentiation formulae (\ref{temp63}) give a very efficient procedure
for the calculation of diagonal correlation functions $C^\pm(N,N)$
for the symmetric Ising model.

\section{A calculation of $C^\pm(0,1;\l)$}

In this section we perform a resummation of
the $\l$-extension
of the simplest row nearest-neighbor correlation
$C^\pm(0,1;\l)$.
It is not known whether off-diagonal correlations
and their $\l$-extensions
satisfy nonlinear differential equations.
Therefore, a knowledge of explicit expressions
for $C^\pm(M,N;\l)$ at
small values of $M$, $N$ can be useful.

First we shall start with the expansion of
integrals (\ref{temp5}) at $s\to0$.
We have
\bea
&&\sinh\g_i=s^{-1}(1-s\cos\omega_i+O(s^2)),\quad
x_i=\frac{s}{2}(1+s\cos\omega_i+O(s^2)),\nonumber\\
&&h_{ij}^2=\sin^2\left(\frac{\omega_i-\omega_j}{2}\right)\,s^2\,(
1+s(\cos\omega_i+\cos\omega_j)+O(s^2)).\label{temp75}
\eea
Then it is easy to see that  $\hat C^{(n)}(0,1)$ has a kinematic
zero of order $n^2$ as $s\to0$.

Expanding $\hat C^{(n)}(0,1)$ in $s$ for $n=1,2,3,4,5,6$
one can see that $\hat C^{(n)}(0,1)$ has a zero in $s$ of order
$n(n+1)$ and guess the
first
two nonzero terms of the expansion
\beq
\hat C^{(n)}(0,1)=
\frac{s^{n(n+1)}}{2^{n^2}}(1+\frac{s^2}{4}+O(s^4))
\quad \mbox{at}\quad s\to 0,\quad n=1,2,3,\ldots\label{temp76}
\eeq

We expect these integrals to be polynomials in
$K$, $E$ and $s^2$ \cite{Holon}.
The case $n=1$ is easy
\beq
\hat C^{(1)}(0,1)=-\frac{1}{2}+\frac{1}{2}(1+s^2)K,\label{temp77}
\eeq
and the answer for $n=2$ was given in \cite{Holon}
\beq
\hat C^{(2)}(0,1)=\frac{3}{8}-\frac{1}{4}(1+s^2)K-\frac{1}{8}
(s^2-3)(1+s^2)K^2-\frac{1}{2}E K.\label{temp78}
\eeq
Using series expansions in $s$ up to a sufficiently high order
we were able to obtain similar expressions for $n=3,4$.
They are given in Appendix B.
The structure of the $n=1,2,3,4$ cases
suggests the following Ansatz
for general $n:$
\beq
\hat C^{(n)}(0,1)=\sum_{i=0}^{[\frac{n}{2}]}a^{(n)}_{i}(E K)^i+
(1+s^2)\sum_{i=1}^n
\sum_{j=0}^{[\frac{i-1}{2}]}b^{(n)}_{i,j}E^jK^{i-j}p_{i-2j}(s)
\label{temp79}
\eeq
where $[x]$ stands for the integer part of $x$,
$a^{(n)}_{i}$ and $b^{(n)}_{i,j}$ are rational coefficients,
and $p_n(s)$ is the polynomial in $s^2$ of degree $n-1$
with integer coefficients:
\beq
p_n(s)=c_{0,n}+c_{1,n}s^2+\ldots+
c_{n-2,n}s^{2(n-2)}+s^{2(n-1)}.\label{temp80}
\eeq

It is quite remarkable  that the condition for the sum in
(\ref{temp79}) to have a zero of order $n(n+1)$ in
$s$  fixes all coefficients in (\ref{temp79}) together
with polynomials $p_n(s)$ up to a common normalization
factor. This factor
is determined by the coefficient of the leading term
in (\ref{temp76}).

Applying this procedure  we easily calculated
$\hat C^{(n)}(0,1)$ for $n=5,6,\ldots,15$.
The coefficients $a^{(n)}_{i}$ and $b^{(n)}_{i,j}$
together with polynomials $p_n(s)$ up to $n=6$ are
listed in Appendix B.

Of course, this is just experimental mathematics based on matching
series expansions. It is desirable to run
some self-consistency tests.
We evaluated the integrals $\hat C^{(n)}(0,1)$ for $n=1,\ldots,6$
numerically at $s=0.95$ and they matched (\ref{temp79}) with 1 part in
$10^{-8}$ accuracy which corresponds to approximately $ 360$ terms
in the series expansions in $s$. So the Ansatz (\ref{temp79})
is almost certainly correct. Note that eqn (5.8) in Orrick et al.
\cite{ONGP00} gives a conjectured expression for $\hat C^{(n)}(0,1)$
in terms of Jacobi theta functions and $q$ series. Expanding that
equation in a $q$-series and converting the series expansions of
(B1) to (B7) to $q$-series provides a valuable consistency check.
We confirm that the series agree to the first 100 terms.

Now having explicit expressions for the form-factors
we can try to perform
a resummation of the next-neighbor correlations $C(0,1;\l)$.

As we did for the diagonal correlations, it is
natural to look at the
the logarithms of $C^\pm(0,1;\l)$ and expand
them in series in $x,$ where
\beq
x=\arcsin(\l).\label{temp81}
\eeq
Then we obtain
\beq
\log C^\pm(0,1;x)=
-\frac{K E}{2}\,x^2+f^\pm(x,K,s^2)\label{temp82}
\eeq
where functions $f^\pm(x,K,s^2)$ as series in $x$ do not depend
on the elliptic integral of the second kind $E$.
The absence of $E$ in the functions $f^\pm(x,K,s^2)$ is a very strong
argument in favour of our Ansatz  (\ref{temp79}).

Using definitions (\ref{temp3}-\ref{temp4}) and (\ref{temp51}) we
can write
\beq
C^+(0,1;x)=\frac{1}{s}\frac{\theta_4(x|\tau)}{\theta_3(0|\tau)}\,
F^+(x,K,s^2),\quad
C^-(0,1;x)=\frac{\theta_4(x|\tau)}{\theta_3(0|\tau)}\,
F^-(x,K,s^2).\label{temp83}
\eeq
Now let us rewrite series expansions for
$F^\pm(x,K,s^2)$ in variables
$x$ and $y=xK$ and split them into even
and odd parts in  $y$.
It appears that the resulting two functions factorize
and have similar structure
\beq
F^\pm(x,K,s^2)=
\a_\pm(x)\left[g_{even}(xK,s^2)-
\b_\pm(x)\,g_{odd}(xK,s^2)\right].\label{temp84}
\eeq
where $\a_\pm(x)$ and $\b_\pm(x)$ depend on
the variable $x$ only
and $g_{even}(y,s^2)$, $g_{odd}(y,s^2)$ do
not depend on $x$.
A similar separation happens for generalized
diagonal correlations
at $N=0,1$ and is another strong argument in
favour of our calculations.

Using Ansatz (\ref{temp79}) we calculated  form factors (\ref{temp5})
for $n=1,\ldots,15$ and
obtained series expansions for  functions
$\a_\pm(x)$ and $\b_\pm(x)$ up to
$x^{15}$. Then it is easy to guess that
\beq
\a_+(x)=-\frac{\sin\bigl(\frac{x}{2}\bigr)}{\cos(x)},\quad
\a_-(x)=\frac{\cos\bigl(\frac{x}{2}\bigr)}{\cos(x)},\quad
\b_-(x)=\frac{1}{\b_+(x)}=\tan\Bigl(\frac{x}{2}\Bigr).\label{temp85}
\eeq
Finally we have to find the functions
$g_{even}(y,s^2)$ and $g_{odd}(y,s^2)$
where the variable $y=xK$.
Their expansions in $y$ have polynomial
coefficients in $s^2$ of growing
complexity and it is not easy to reconstruct them.

However, based on the diagonal case we assumed
that these functions are double-periodic
in $y$ and substituted their expansions into the differential equation
\beq
(\partial_y g(y,s))^2=\sum_{i=0}^4 A_i(s)(g(y,s))^i,\label{temp86}
\eeq
where the coefficients $A_i(s)$ are independent of the variable $y$.

This leads to immediate success, and we can
easily find the polynomial coefficients $A_i(s)$
and solve the differential equation (\ref{temp86}) for both cases.

Collecting all the above formulae we get the final answer
\begin{eqnarray}
C^-(0,1;x)=\phantom{\frac{1}{2s}}\frac{\theta_4(x|q)}{\theta_3(0|q)}
\frac{\cos\bigl(\frac{x}{2}\bigr)}{\cos(x)}
\left[g_{ev}(x,s)-
\tan\Bigl(\frac{x}{2}\Bigr)\,g_{odd}(x,s)\right],&&\nonumber\\
C^+(0,1;x)=
-\frac{1}{s}\frac{\theta_4(x|q)}{\theta_3(0|q)}\frac{
\sin\bigl(\frac{x}{2}\bigr)}{\cos(x)}
\left[g_{ev}(x,s)-
\cot\Bigl(\frac{x}{2}\Bigr)\,g_{odd}(x,s)\right],&&\label{temp87}
\end{eqnarray}
where
\begin{equation}
g_{ev}(x,s)=\frac{\mbox{cn}(\frac{xK}{2},s^4)\mbox{dn}(\frac{xK}{2},s^4)}
{1-s^2\mbox{sn}(\frac{xK}{2},s^4)^2},\quad
g_{odd}(x,s)=\frac{(1+s^2)\,\mbox{sn}(\frac{xK}{2},s^4)}
{1+s^2\mbox{sn}(\frac{xK}{2},s^4)^2}\label{temp88}
\end{equation}
and the variable $x$ is related to $\l$ by
(\ref{temp54}) or (\ref{temp81}).

The final check is  to expand (\ref{temp87})
near $x=\pi/2$
which corresponds to the symmetric Ising model
and compare the answer
with known nearest-neighbor row pair correlation
functions. We obtain
\beq
C^+(0,1)=\frac{\sqrt{1+s^2}}{2s}(1+(s^2-1)K),\label{temp89a}
\eeq
and
\beq
C^-(0,1)=\frac{\sqrt{1+s^2}}{2}(1-(s^2-1)K),\label{temp89}
\eeq
which match (4.5a) and (4.5b) of
Section VIII.4, \cite{McCoyWu}.

Let us briefly comment on the formulae (\ref{temp87}).
To our knowledge closed expressions for generalized diagonal
correlations $C^\pm(0,1)$ have not appeared in the literature before.

As discussed in in the introduction in 1980 McCoy, Perk and Wu 
obtained quadratic difference
relations for correlation functions. In our notation
the $\l$-extension of
these  equations  is given in \cite{ONGP00}.
These difference equations allow one to calculate generalized correlation
functions provided that we know the answer for diagonal correlations
{\it and} the first off-diagonal pair correlations $C(0,1)$.
Therefore, formulae (\ref{temp87}) provide the
initial conditions for a recursive calculation of all generalized
pair correlation functions $C(M,N;\l)$ for arbitrary $M,N=0,1,2,\ldots$

While undoubtedly correct, a rigorous proof is still lacking.
The missing ingredient is, for example, a Toda type relation similar
to that obtained above for the diagonal correlations,
plus verification of the connection between the parameters $\l$ and $x$.
This is currently under investigation.

After the appearance of the first draft of this paper,
it was pointed out to us by W.P. Orrick and J-M. Maillard  that
the result we give in eqn (\ref{temp87}) is, in principle, derivable
from eqn (5.8) in the paper \cite{ONGP00} by Orrick et al. by summing
over the correlation functions and making appropriate substitutions.
We have not confirmed this result. We remark however that if that is
indeed the case, further exact expressions for $ C(1,1;\lambda),$
$C(2,0;\lambda),$ and $C(2,1;\lambda)$ should be similarly
derivable from eqns (5.9)--(5.11) of \cite{ONGP00} and consistent
with Perk's difference equations.

\section{Conclusions and outlook}

In this paper we have found and proved a set of Toda-type difference
equations (\ref{intro6}) satisfied by diagonal pair correlation functions.
It is interesting that the $\l$- extension of these pair correlations
satisfy the same equations. We think that there should be a direct way
to prove relations (\ref{intro6})
from the Toeplitz determinant representation of the diagonal correlations.
That would then enable us to understand the determinantal representation
for their $\l$--extension.

When that is done, we could try to generalize relations (\ref{intro6})
for the case of row correlations where the Toeplitz determinant
representation is known for $\l=1$. However,
it could well be difficult to generalize it to arbitrary values of $\l$.
In fact, formulae (\ref{temp87}) provide the first examples of closed form expressions
for  the $\l$- extended nearest-neighbor row correlation functions.
We failed to find a nonlinear differential equation satisfied
by (\ref{temp87}) valid for {\it any value of $\l$}.
Of course, that doesn't mean that such an equation doesn't exist.

Let us emphasize that based on connections to Painlev\'e VI theory
we have proved the results in equations (\ref{temp50}, \ref{temp58}),
up to the non-rigorous identification of the connection between
the parameter $\lambda$ and the parameters $x$ and $y.$

We also note that in 2005 Forrester and Witte \cite{FW05} gave
a nonlinear system of recurrences in their corollary (6.3), related to the
discrete Painlev\'e V equation for the same
ratio of correlations that appears on the rhs of  (\ref{temp38}).

Finally, as we mentioned above, the derivation of the generalized
diagonal correlations
and relations (\ref{temp87}) is not rigorous, based as it is on
series expansions. However it is undoubtedly correct, satisfying
numerous checks, and should provide  initial
conditions to calculate all generalized pair correlations
$C(M,N;\l)$
for any $M$ and $N$ from the quadratic difference
equations of Perk. This should provide all necessary
tools for a significantly more efficient generation
of the series expansion for the $\l$-extended
susceptibility. An ambitious, but still distant hope is
to sum the correlation function series and produce
the long sought closed-form expression for the
susceptibility. We are planning to report on the
evaluation of $C(M,N;\l)$ in a
forthcoming publication.

\section*{Acknowledgments}

The authors thank V.V. Bazhanov, M.T. Batchelor, J. de Gier
and  S. Sergeev
for valuable comments and interesting discussions while this
work was being conducted, and P. J. Forrester, B.G Nickel,
J-M Maillard, B M Mc Coy, B.G Nickel and W P Orrick for
valuable comments which significantly improved the manuscript. This work is
supported by the Australian Research Council.

\app{The diagonal case}

In this Appendix we present the first seven form factors (\ref{temp38})
at $M=N=0$ used for the calculation of initial conditions for the
recursion relations (\ref{intro6}). We also give generalized
diagonal correlation functions for $N=3$.

Form factors $\hat C^{(n)}(0,0)$ for $n=1,\ldots,7:$
\bea
&& \frac{\hat C^{(1)}(0,0)}{sK}= 1,\nonumber\\\nonumber \\
&& \frac{\hat C^{(2)}(0,0)}{K}=
\frac{K}{2}-\frac{E}{2},\nonumber\\\nonumber \\
&& \frac{\hat C^{(3)}(0,0)}{sK}=\frac{1}{6}-
\frac{1}{2}KE-\frac{1}{6}(t-2)K^2,\nonumber\\\nonumber \\
&& \frac{\hat C^{(4)}(0,0)}{K}=
\frac{K}{6}+\frac{K^3}{8}-\frac{t\,K^3}{12}-
\frac{E}{6}-\frac{K^2E}{4}+\frac{KE^2}{8}\nonumber\\\
\nonumber \\
&&  \frac{\hat C^{(5)}(0,0)}{sK}=\frac{3}{40}-\frac{KE}{4}+
\frac{K^2E^2}{8}
-\frac{1}{12}(t-2)K^2(1-KE)+\frac{1}{120}(t^2-6t+6)K^4.
\eea
\bea
&&  \frac{\hat C^{(6)}(0,0)}{K}=\frac{4K}{45}+
\frac{K^3}{12}\left(1-\frac{2t}{3}\right)+
\frac{K^5}{48}\left(1-\frac{22t}{15}+\frac{8t^2}{15}\right)+
\nonumber\\\
&&{\phantom{ \frac{\hat C^{(6)}(0,0)}{sK}}}+E
\left(-\frac{4}{45}+\frac{KE}{12}-\frac{K^2E^2}{48}-
\frac{K^2}{6}+\frac{K^3E}{16}+\frac{t\,K^4}{24}-\frac{K^4}{16}
\right)\nonumber \\ \nonumber\\
&&\frac{\hat C^{(7)}(0,0)}{sK}=\frac{5}{112}-\frac{37KE}{240}+
\frac{5K^2E^2}{48}-\frac{K^3E^3}{48}
-(t-2)K^2\left(\frac{37}{720}-\frac{5KE}{72}+\frac{K^2E^2}{48}\right)+
\nonumber\\
&&\phantom{\frac{\hat C^{(1)}(0,0)}{sK}}+(t^2-6t+6)K^4
\left(\frac{1}{144}-\frac{KE}{240}\right)-
\frac{1}{5040}(t-2)(t^2-10t+10)K^6
\eea

In \cite{Bos09} it is pointed out by Bostan et al. that
the minimal order ODE for the $n$-particle contributions
to the susceptibility $\tilde{\chi}^{(n)}$ are achieved
by appropriate linear combinations of $\tilde{\chi}^{(n)}$
with terms $\tilde{\chi}^{(n-2)},$ $\tilde{\chi}^{(n-4)},$ etc.
In particular, these combinations are:
$\tilde{\chi}^{(3)} -\tilde{\chi}^{(1)}/6,$
$\tilde{\chi}^{(4)} -\tilde{\chi}^{(2)}/3,$
$\tilde{\chi}^{(5)} -\tilde{\chi}^{(3)}/2 + \tilde{\chi}^{(1)}/120,$
$\tilde{\chi}^{(6)} -2\tilde{\chi}^{(4)}/3 + \tilde{\chi}^{(2)}/45.$
It is remarkable that the analogous combinations of form factors
$\hat C^{(n)}(0,0)$ for $n=1,\ldots,7$ give expressions that are
homogeneous in products of elliptic integrals of degree
$n.$~\footnote{This was pointed out to us by B~G~Nickel.}
That is to say, combinations such as
$\hat C^{(6)}(0,0) - 2\hat C^{(4)}(0,0)/3 + \hat C^{(2)}(0,0)/45$
contain only terms of the form $K^6,$ $K^5 E,$ $K^4 E^2$
and $K^3 E^3.$ From (A.2) it is easy to conjecture that
the appropriate linear combination that will give
the minimum order ODE for $\tilde{\chi}^{(7)}$ is
$\tilde{\chi}^{(7)} -5\tilde{\chi}^{(5)}/3! +
13\tilde{\chi}^{(3)}/5! - \tilde{\chi}^{(1)}/7! $

Generalized diagonal correlations $C^\pm_3(t)$ are
obtained by solving (\ref{intro6})
for $N=2. $ They are:

\beq
C^+_3(t)=
\frac{\theta_4(x|\tau)}{135\,t^{5/2}\,\cos^3(x)\,\theta_3(0|\tau)}
\Bigl[t\,C\,S\,R_1^++XR_2^+\Bigr],
\eeq
and
\beq
C^-_3(t)=
-\frac{\theta_4(x|\tau)}{135\,t^2\,\cos^3(x)\,\theta_3(0|\tau)}
\Bigl[C\,R_1^-+S\,XR_2^-\Bigr],
\eeq
\bea
&&R_1^+(t)=168\,X^6+84\,X^4\left[6\,t\,S^2-7(t+1)\right]_{\phantom{|}}
+4\,t\,(1-S^2)(1-t\,S^2)(2\,t\,S^2-9(t+1))\nonumber\\
&&+108\,X^2\left[2\,t\,S^2
(t\,S^2-3(t+1))+3+4t+3t^2\right]
,\\
\nonumber\\
&&R_2^+(t)=-X^8+6X^6\left[6\,t\,S^2-t-1\right]+9X^4\left[
42\,t\,S^2(t\,S^2-t-1)-t^2+28\,t-1\right]_{\phantom{|}}\nonumber\\
&&+\,
2X^2\left[105\,t^2S^4(2\,t\,S^2-5(t+1))+3\,t\,S^2(99+196\,t+99\,t^2)-
(t+1)(2+103\,t+2\,t^2)\right]_{\phantom{|}}
\nonumber\\
&&+\,3\,t\,(1-S^2)(1-t\, S^2)\left[7\,t\,S^2(3\,t\,S^2-
11(t+1))+24\,t^2+21\,t+24\right],
\eea
\bea
&&R_1^-(t)=9X^8+42X^6\left[2\,t\,S^2-t-1\right]+3X^4\left[
14\,t\,S^2(3\,t\,S^2-t-1)-13t^2+28\,t-13\right]_{\phantom{|}}
\nonumber\\
&&+\,
6X^2\left[3\,t^2S^4(2\,t\,S^2+t+1)-3\,t\,S^2(3+52t+3t^2)+
(t+1)(2+13t+2t^2)\right]_{\phantom{|}}
\nonumber\\
&&+\,t^2(1-S^2)(1-t\, S^2)\left[t\,S^4+3\,S^2(t+1)-135\right],\\
\nonumber\\
&&R_2^-(t)=X^8+6X^6\left[6\,t\,S^2-5(t+1)\right]+9X^4\left[
14\,t\,S^2(t\,S^2-t-1)+t^2+28\,t+1\right]_{\phantom{|}}\nonumber\\
&&+\,
2X^2\left[21\,t^2S^4(2S^2t-t-1)-3\,t\,S^2(13+84t+13t^2)+
(t+1)(20+61\,t+20\,t^2)\right]_{\phantom{|}}
\nonumber\\
&&+\,3\,t\,(1-S^2)(1-t\, S^2)\left[t\,S^2(3\,t\,S^2+5(t+1))-
4\,t^2-197\,t-4\right].
\eea

\app{The off-diagonal case}
In this Appendix we first give   form factors $\hat C^{(n)}(0,1)$
for $n=1,2,3,4$
obtained by series expansions of (\ref{temp5}) to series in $s$
up to 200 terms.

\bea
\hat C^{(1)}(0,1)&=&-\frac{1}{2}+\frac{1}{2}(1+s^2)K,\\ \\
\hat C^{(2)}(0,1)&=&\phantom{-}\frac{3}{8}-\frac{1}{4}(1+s^2)K-
\frac{1}{8}(s^2-3)(1+s^2)K^2-\frac{1}{2}E K,\\ \\
\hat C^{(3)}(0,1)&=&-\frac{5}{16}+\frac{13}{48}(1+s^2)K+
\frac{1}{16}(s^2-3)(1+s^2)K^2-\\ \nonumber
&& -\frac{1}{48}(s^4+6s^2-11)(1+s^2)K^3+\frac{1}{4}E K-
\frac{1}{4}E K^2(1+s^2),\\ \\
\hat C^{(4)}(0,1)&=&\phantom{-}\frac{35}{128}-
\frac{19}{96}(1+s^2)K-\frac{17}{192}(s^2-3)(1+s^2)K^2+\\ \nonumber
&&+\frac{1}{96}(s^4+6s^2-11)(1+s^2)K^3+
\frac{1}{384}(s^6-21s^4+3s^2+25)(1+s^2)K^4-\\ \nonumber
&&-\frac{17}{48}E K+\frac{1}{8}E^2K^2+
\frac{1}{8}E K^2(1+s^2)+\frac{1}{16}EK^3(s^2-3)(1+s^2)
\eea

We note that, in analogy to the linear combinations of form
factors $\hat C^{(n)}(0,0)$ for $n=1,\ldots,7$ that give
 expressions that are homogeneous in products of elliptic
 integrals of degree $n$ reported in Appendix A, rather
 more complicated combinations of form factors $\hat C^{(n)}(0,1)$
 produce similar results for the form factors $\hat C^{(n)}(0,1).$
 In particular, we note that
$$\hat C^{(3)}(0,1) + \frac{1}{2}\hat C^{(2)}(0,1) -
\frac{7}{24}\hat C^{(1)}(0,1) - \frac{1}{48} = {\rm O}(EK^2, K^3)$$ and
$$\hat C^{(4)}(0,1)+\frac{1}{2}\hat C^{(3)}(0,1) -
\frac{11}{24}\hat C^{(2)}(0,1) - \frac{5}{48}\hat C^{(1)}(0,1) +
\frac{1}{384} = {\rm O}(EK^3, E^2K^2, K^4).
$$
Here we also list coefficients
$a^{(n)}_{i}$ , $b^{(n)}_{i,j}$ and polynomials
$p_n(s)$ for $n=1,\ldots,6$ which occur in  eqn (\ref{temp79})
for the form factors $\hat C^{(n)}(0,1)$.

%\suppressfloats
\begin{table}
\begin{center}
$
\begin{array}{|r|r|r|r|r|}
\hline
n&{a_{0}^{(n)}}^
{\hspace{-0.2cm}\phantom{|}}_{\hspace{-0.2cm}\phantom{|}}&
a_{1}^{(n)}&a_{2}^{(n)}&a_{3}^{(n)}\\\hline
6&{\frac{231}{1024}}^
{\hspace{-0.2cm}\phantom{|}}_{\hspace{-0.2cm}
\phantom{|}}& -\frac{3319}{11520}& \frac{25}{192}
&-\frac{1}{48}\\\hline 5&{-\frac{63}{256}}^
{\hspace{-0.2cm}\phantom{|}}_{\hspace{-0.2cm}
\phantom{|}}& \frac{23}{96}& -\frac{1}{16}\\\cline{1-4}
4&{\frac{35}{128}}^
{\hspace{-0.2cm}\phantom{|}}_{\hspace{-0.2cm}
\phantom{|}}& -\frac{17}{48}& \frac{1}{8}\\\cline{1-4}
3&{-\frac{5}{16}}^{\hspace{-0.2cm}
\phantom{|}}_{\hspace{-0.2cm}\phantom{|}}& \frac{1}{4}\\
\cline{1-3}
2&{\frac{3}{8}}^{\hspace{-0.2cm}\phantom{|}}_{
\hspace{-0.2cm}\phantom{|}}& -\frac{1}{2}\\
\cline{1-3}
1&{-\frac{1}{2}}^{\hspace{-0.2cm}\phantom{|}}_{
\hspace{-0.2cm}\phantom{|}}\\
\cline{1-2}
\end{array}$
\end{center}
\caption{Coefficients $a_{i}^{(n)}$ for $n=1,\ldots,6.$}
\end{table}

\begin{table}
$
\begin{array}{|r|r|r|r|r|r|r|r|r|r|r|r|r|}
\hline
n&{b_{1,0}^{(n)}}^{\hspace{-0.2cm}\phantom{|}}_{
\hspace{-0.2cm}\phantom{|}}& b_{2,0}^{(n)}&
b_{3,0}^{(n)}&b_{3,1}^{(n)}&
b_{4,0}^{(n)}&b_{4,1}^{(n)}&
b_{5,0}^{(n)}&b_{5,1}^{(n)}&b_{5,2}^{(n)}&b_{6,0}^{(n)}&
b_{6,1}^{(n)}&b_{6,2}^{(n)}\\
\hline
6&-{\frac{1289}{7680}}^{\hspace{-0.2cm}\phantom{|}}_{
\hspace{-0.2cm}\phantom{|}}& -\frac{3319}{46080}& \frac{3}{256}&
\frac{9}{64}&\frac{25}{9216}&\frac{25}{384}&
-\frac{1}{7680}&-\frac{1}{192}&-\frac{1}{32}&-\frac{1}{46080}&
-\frac{1}{768}&-\frac{1}{64}\\
\hline
5&{\frac{263}{1280}}^{\hspace{-0.2cm}\phantom{|}}_{
\hspace{-0.2cm}\phantom{|}}& \frac{23}{384}& -\frac{7}{384}&
-\frac{7}{32}&-\frac{1}{768}&-\frac{1}{32}&\frac{1}{3840}&
\frac{1}{96}& \frac{1}{16}\\
\cline{1-10}
4&-{\frac{19}{96}}^{\hspace{-0.2cm}\phantom{|}}_{
\hspace{-0.2cm}\phantom{|}}& -\frac{17}{192}& \frac{1}{96}&
\frac{1}{8}&\frac{1}{384}&\frac{1}{16}\\
\cline{1-7}
3&{\frac{13}{48}}^{\hspace{-0.2cm}\phantom{|}}_{
\hspace{-0.2cm}\phantom{|}}& \frac{1}{16}
&-\frac{1}{48}&-\frac{1}{4}\\
\cline{1-5}
2&-{\frac{1}{4}}^{\hspace{-0.2cm}\phantom{|}}_{
\hspace{-0.2cm}\phantom{|}}& -\frac{1}{8}\\
\cline{1-3}
1&{\frac{1}{2}}^{\hspace{-0.2cm}\phantom{|}}_{
\hspace{-0.2cm}\phantom{|}}\\
\cline{1-2}
\end{array}\label{tab2}
$
\caption{Coefficients $b_{i,j}^{(n)}$ for $n=1,\ldots,6$}
\end{table}

\bea
&&p_1(s)=1,\nonumber\\
&&p_2(s)=-3+s^2,\nonumber\\
&&p_3(s)=-11+6s^2+s^4,\nonumber\\
&&p_4(s)=25+3s^2-21s^4+s^6,\nonumber\\
&&p_5(s)=201-180s^2-66s^4+60s^6+s^8,\nonumber\\
&&p_6(s)=-299-123s^2+466s^4+106s^6-183s^8+s^{10}.
\eea
\newpage
%\bibliographystyle{vvb-bibstyle}

%\bibliography{refs225}

%\newcommand\oneletter[1]{#1}

\end{document}